\documentclass{article}

\usepackage{microtype}
\usepackage{graphicx}
\usepackage{subcaption}
\usepackage{booktabs}
\usepackage{eurosym}
\PassOptionsToPackage{hyphens}{url}
\usepackage{hyperref}
\usepackage{xurl}
\usepackage{amsmath,amssymb,mathtools,amsthm}
\usepackage{algorithm}
\usepackage{algorithmic}
\usepackage[capitalize,noabbrev]{cleveref}

\usepackage[accepted]{icml2026}


\icmltitlerunning{Fingerprinting AI Cluster I/O Without Trusted Processors}

\begin{document}

\twocolumn[
\icmltitle{Fingerprinting All AI Cluster I/O\\
           Without Mutually Trusted Processors}

\icmlsetsymbol{equal}{*}

\begin{icmlauthorlist}
\icmlauthor{Naci Cankaya}{mats,haigl}
\icmlauthor{Jakub Kry\'s}{saferai}
\icmlauthor{Jonathan Ng}{ind}
\icmlauthor{Luke Marks}{martian}
\icmlauthor{Felix Krueckel}{rwth}
\end{icmlauthorlist}

\icmlaffiliation{mats}{MATS}
\icmlaffiliation{haigl}{Oxford Hardware AI Governance Lab}
\icmlaffiliation{saferai}{SaferAI}
\icmlaffiliation{ind}{Independent}
\icmlaffiliation{martian}{Martian}
\icmlaffiliation{rwth}{RWTH Aachen University, Aachen, Germany}

\icmlcorrespondingauthor{Naci Cankaya}{naci.c@protonmail.com}

\icmlkeywords{AI governance, verification, covert channels,
              steganography, active warden, data centre, FPGA,
              compute governance}

\vskip 0.3in
]

\printAffiliationsAndNotice{}

\begin{abstract}
In preparation for potential international agreements on artificial intelligence, the development of verification infrastructure for AI data centres is vital. We propose a method for cryptographically committing all information entering and leaving a data centre: Hashes are computed by network taps placed on all the information-carrying wires between the cluster and the outside world, enabling an auditor to retroactively challenge the preimage data to be sent to a privacy-preserving verification facility performing compliance checks. Our goal is to make it infeasible to covertly exfiltrate the results of undisclosed workloads in the cluster through the tapped wires. To this end, we specify the architecture of a ``Secure Gateway Device'', which handles the erasure of covert channels that post-hoc verification on hashed data cannot address: analogue and timing side-channels, as well as steganography in network protocol headers. The architecture eliminates the need for any processors trusted by both the Prover and the Verifier, leveraging passive optical fibre splitters and coin-flip protocols for random number generation where needed. We expect development costs of a demonstration device to be roughly equivalent to the cost of a small team of engineers for a few months, with a comparatively small bill of materials.
\end{abstract}

\section{Introduction}
\label{sec:intro}

The rapid progress of Artificial Intelligence (AI) in recent years has opened the door to accelerating advancements in multiple fields including medicine, engineering and computer science. However, as a very powerful general-purpose tool, AI also exacerbates existing concerns around technology misuse, as well as introduces novel risk vectors such as the threat of autonomous misaligned agents~\cite{bengio2026safetyreport}. For this reason, several jurisdictions have already introduced legally-binding regulations governing the acceptable uses, transparency obligations or testing requirements of generative AI systems~\cite{cac2023interim,eu2024aiact,wiener2025sb53}. Additionally, in the future the need and political will for treaties and agreements might also arise between different jurisdictions. For example, in recognition of the tremendous power afforded by AI, leading actors in this technology competition could agree on `red lines' that neither of them should cross~\cite{cesia2025redlines}. These may emulate historical agreements in other safety-critical areas such as The Nuclear Non-Proliferation Treaty or START.

Regardless of the nature of such legislation or treaties, there is a clear need for the ability to verify compliance with agreements on AI by all affected parties. This problem is particularly important and challenging in the context of AI, since computing power as a resource has a much more dual-use nature than fissile materials or even DNA synthesis tools~\cite{sastry2024compute}. Advanced artificial intelligence is seen as a decisive strategic asset and economic engine~\cite{nscai2021final}, and the question of whether unilateral restraint in AI development and deployment is compatible with national security is contested.\footnote{US Vice President Vance, when asked about pausing dangerous AI development: ``The honest answer to that is that I don't know, because part of this arms race component is if we take a pause, does the People's Republic of China not take a pause? And then we find ourselves all enslaved to P.R.C.-mediated A.I.?''~\cite{douthat2025vance}}

\begin{figure}[htbp]
  \centering
  \includegraphics[width=\columnwidth]{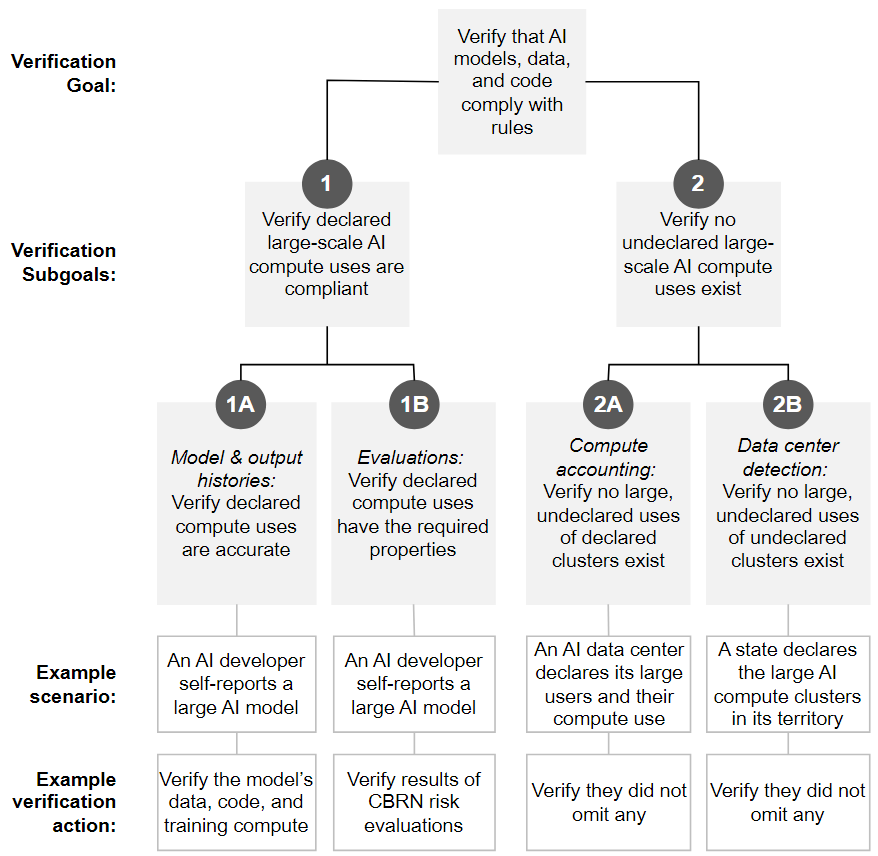}
  \caption{A breakdown of the AI Verification research agenda into its goals and subgoals. Original from \citet{baker2025verifying} with the authors' permission. In this work, we focus on verification Subgoals 1A and 2A.}
  \label{fig:agenda}
\end{figure}

Consequently, the field of AI Verification has recently emerged as a research agenda spanning both technical and governance aspects. The importance of robust verification mechanisms is reflected by the appearance in literature of three major works detailing open problems, policy goals and possible research directions in this area~\cite{scher2024mechanisms,harack2025verification,baker2025verifying}. These are presented concisely in \Cref{fig:agenda}.

In this work, our main focus are Subgoals 1A and 2A of \citet{baker2025verifying}, which can be summarised as follows: an entity operating a computing cluster --- the Prover --- aims to prove to the auditing party --- the Verifier --- that no undeclared workloads were executed on said cluster. The Verifier checks if the evidence was generated in a way the Prover can not tamper with, and if the claims are self-consistent/accurate.

We do not engage with the problem of determining which workloads should be allowed or not in the first place (regulation) or of verifying whether the declared workloads are compliant (Subgoal~1B). However, our proposed method offers raw evidence of compute use, which is needed for the downstream verification of correctness (Subgoal~1A) and compliance (Subgoal~1B). The issue of \textit{preventing undeclared compute resources} (Subgoal~2B) is fully outside of the scope of this work.

Proposed solutions for verifying compliance and comprehensiveness include on‑chip mechanisms such as Trusted Execution Environments, HBM‑based chiplets that monitor memory traffic~\cite{petrie2025guaranteeable, aarne2024chips}, and tamper‑proof enclosures for accelerators~\cite{happel2026tampersec}.\footnote{For a thorough overview of possible directions, we refer the reader to the three‑part series by the FlexHEG initiative~\cite{petrie2025flexheg}.} Off‑chip, network‑level approaches have also been described: \citet{baker2025verifying} detail an explicit network‑tap layer, \citet{harack2025verification} introduce ``digital perimeters'' that commit to cluster traffic, and \citet{scher2024mechanisms} explore how external networking equipment can enforce workload‑classifying bandwidth limits.  We build directly on these off‑chip paradigms.  Instead of targeting individual accelerators or servers, we operate at the level of the whole cluster, focusing on the communication chokepoint between the hardware and the Prover---the data centre north‑south uplink (connecting the data centre's clusters with the internet/outside world). If the Verifier can capture, hash and timestamp all traffic in and out of the data centre, and the Prover can --- upon later request --- demonstrate that their declared workloads correspond to this hashed traffic, the Verifier can be satisfied that the Prover is not performing undeclared computations in the cluster. To achieve this goal, we propose a schematic design of a ``Secure Gateway Device'' (SeGaDev) that uses active and passive network taps to capture and process such traffic for later verification in a mutually trusted manner.

\paragraph{Main contributions.}
\begin{itemize}
\item A detailed description of the problem statement and the design requirement for a verification mechanism meant to address Subgoal 2A of \citet{baker2025verifying} (\Cref{sec:problem}).
\item An actionable design specification of our proposed device that addresses these requirements. This includes a detailed description of how the device processes captured traffic in order to eliminate steganographic degrees of freedom in egress data (\Cref{sec:architecture}).
\item A feasibility assessment of this solution, including estimating the storage requirements needed for the associated verification protocol (\Cref{sec:covert} and \Cref{app:storage}).
\end{itemize}

\section{Background and Related Work}
\label{sec:background}

The security problem of covert communication through an observed link is an active area of research. We differentiate two attack types: side-channels and steganography.

The first exploits physical channels other than the ones used regularly by a device. For example, while an optical fibre link may be specified to only use amplitude modulation for communication, additional information can be encoded in polarization or phase, if the outside observer is not considering them. One physical side-channel that is not trivially closed by active network taps is timing modulation, which is explored by \citet{lee2014phy} on the offensive side and \citet{uttarwar2025delay} on the defensive side. We address this side channel in our proposed solution.

The second channel, steganography, does not hide any signal directly from the observer. Rather, it encodes hidden messages in plaintext. An intuitive example for this is hiding messages in the second character of every word in a text. Applied to Ethernet links (which are the practically universal standard for internet communication), steganography can operate on multiple levels. ~\citet{jankowski2013padsteg} demonstrate this at the protocol layer, encoding secret messages in padding bytes (usually only used to ensure a minimum length of network packets).

To defend against both channels, so-called ``Active Wardens'' were first conceptualized \citep{simmons1984prisoners, anderson1998limits}, prototyped in software \citep{fisk2003eliminating} and, more recently, built with re-purposed network switching hardware \citep{xing2020netwarden}. Active Wardens are devices that intercept and re-emit traffic (closing channel 1), while re-writing free fields (closing channel 2 for those free fields). We draw from this research on Active Wardens for erasing steganography channels in Ethernet protocol headers, as well as work from ~\citet{uttarwar2025delay} for erasing timing side-channels.

While ensuring that computations can be faithfully replayed and verified is not part of our project scope, we refer to~\citet{karvonen2025difr} and ~\citet{cankaya2025catching1} for highlighting the importance of this downstream challenge for closing off exploitable degrees of freedom~\cite{rinberg2025verifying} in the non-deterministic outputs of ML computations.

We discuss our estimates for covert bandwidths in \Cref{sec:covert}.

\section{Problem Statement}
\label{sec:problem}

For our proposed device, we set the following requirements:

\begin{itemize}
\item The device needs to capture all of the ingress/egress traffic of a data centre. This means it is to be placed at the north-south part of the data centre network, intercepting every link with the outside world.\footnote{Strictly speaking, one can parallelize this task across many SeGaDevs. This is an engineering decision with potential tradeoffs regarding the number of tapped links per device.}

\item It must be mutually trusted and/or monitorable by both Prover and the Verifier. This means that:
\begin{itemize}
\item The Prover must not have the ability to hide significant amounts of information in the traffic being monitored. Otherwise, they could transfer undeclared workloads out of the data centre using steganographic techniques, side-channels, timing attacks, and similar.
\item The Verifier must not have the ability to extract Prover's proprietary information out of the device. Concretely, the Verifier receives hashes, and the Prover needs confirmation that these are computed and transmitted correctly.
\end{itemize}

\item The solution should be retrofittable to realistic, full-scale deployments at the scale of $O(100\text{k})$ accelerators and beyond. To the extent possible given the other requirements, it should rely on existing and affordable technologies.

\item The device should introduce as little additional latency as possible in order to minimise the slowdown of usual data centre operations. It should be able to process millions of communication packets streaming in/out every second.\footnote{This is the minimum required to handle a single 100G link, which is a typical bandwidth used at a data centre edge. Though as noted in the previous footnote, one SeGaDev can handle multiple links if adequately equipped.}
\end{itemize}

We are working under the following assumptions:

\begin{itemize}
\item We focus on traffic serving LLM inference API endpoints to external customers. We also assume API traffic egress is at least on the same order of magnitude as ingress, if not larger. This is because customer traffic served via the `streaming API' format, commonly used for chatbot applications, is optimised for low latency in delivering responses to customers at the cost of high data inefficiency. Every single output token is wrapped in a JSON packaging, adding $\sim$200 bytes of overhead.~\cite{openai_streaming_chunks}\footnote{Strictly speaking, streaming responses may contain batches of a few tokens, although in practice the number is often one. We take this simplifying assumption because we are also interested in a worst-case feasibility assessment of the SeGaDev solution. With only 1 token per streaming response, the amount of data produced represents an upper bound for the storage and bandwidth requirements of our method.}

\item Steganography via input payload data (i.e., the plaintext requests to the LLM) is unconstrained.\footnote{The Warden device could in principle rewrite prompts and input media files while preserving semantics. However, this carries complications related to a potential degradation of the inference service quality. We leave this open for future research.} User and developer inputs are free-form and inherently high-entropy.

\item The Prover needs to store all information that forms a part of the hash for as long as the agreement requires, or as long as the Prover wants to be able to demonstrate compliance with rules. In practice, this includes everything the SeGaDev is configured to hash. Our working assumption is a full capture of the payload data (Ethernet protocol layer 7), which includes everything but the connection metadata (which is actively sanitised by the Active Warden part of the SeGaDev).
\end{itemize}

\begin{figure}
  \centering
  \includegraphics[width=\columnwidth]{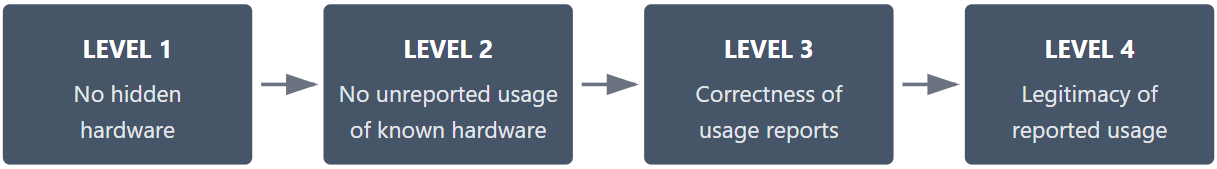}
  \caption{We are specifically aiming for Level 2: capturing and safely transmitting raw evidence of computation (only hashes, not plaintext). The hashes, together with the plaintext data they point to, later feed into a secure cluster trusted by the Verifier, without the plaintext information being shown to them directly.}
  \label{fig:levels}
\end{figure}

\paragraph{Out of scope.}
In order to convey the nuance of the problems we are --- and are not --- tackling, we briefly explain the ``verification pipeline'' we are assuming. This pipeline consists of four separate stages outlined in \Cref{fig:levels}.

The verification of original data in a dedicated, separate cluster is twofold and out of scope for us: (1) replaying observed inputs and outputs to confirm they match precisely (Level 3), and (2) evaluating the computations for legal compliance (Level 4, corresponding to Subgoal~1B in~\citet{baker2025verifying}).\footnote{There are challenges with precise replay. See \Cref{sec:covert}.} While the specific replay protocol is out of scope here, we refer to related work on confidential-computing facilities where the Prover replays challenged computations under attestation and governed egress~\cite{tlaie2026securing}. The Verifier can challenge stored hashes at arbitrary cadence and need not challenge every record.\footnote{ $k$ randomly sampled hashes detect a fraction $p_\text{false}$ of misreported records with probability $1 - (1 - p_\text{false})^k$, independent of total throughput. For instance, $k = 3{,}200$ challenges detect $0.1\%$ misreporting at $96\%$ confidence.}

We do not attempt to solve the physical monitoring of either the Prover's cluster or the verification facility needed to prevent secret communication devices. Also, we do not consider steganographic degrees of freedom in unverifiable information contained in packet payloads (e.g.\ a limit on GPU temperature diagnostics leaving the facility). These degrees of freedom are not for our SeGaDev to erase, but for rules to constrain.\footnote{Since this information is hashed, compliance checking here is trivial: checking if the egress data contains illicit amounts of unverifiable data such as precise timestamps, nonces, GPU temperature and fan speed, power supply metrics etc.}

We only present a network tap at the north-south/frontend network. This does not (directly) prevent covert workloads, but rather the exfiltration of their results. Additional network taps in the east-west/backend network can offer defense-in-depth by directly capturing tensors used in computation. We detail their feasibility and security properties in forthcoming work.

\section{Architecture}
\label{sec:architecture}

In this section, we describe the placement and design of the SeGaDev --- our evidence capture device.

\begin{figure*}[h]
  \centering
  \includegraphics[width=\textwidth]{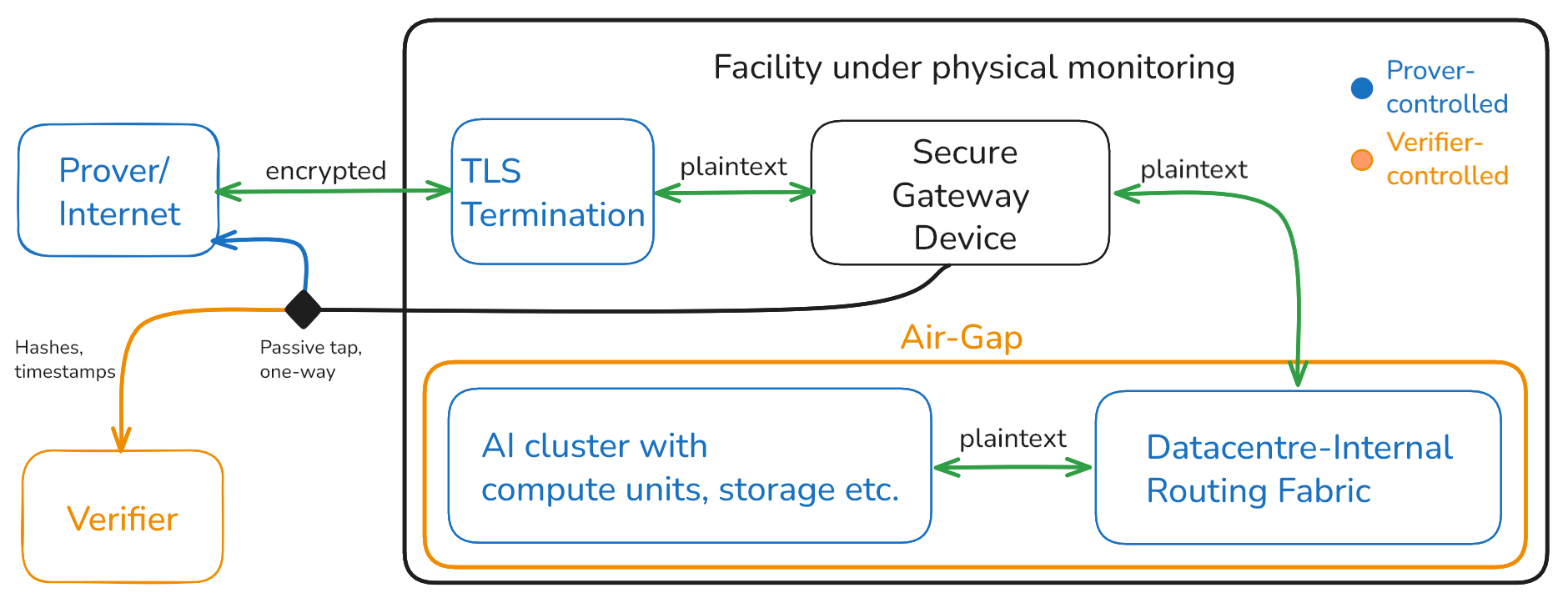}
  \caption{The placement and main task of the SeGaDev. Here, it is shown as an interceptor between TLS termination point (encryption boundary) and the data centre fabric routing traffic to individual compute units. This placement is not the only option: one could also place network taps between the edge router(s) and the compute units, at any level of the north-south network hierarchy. The general principle is that all accelerators are air-gapped, apart from links eventually passing through the SeGaDev. Note the passive splitter at the link carrying hashes to the Verifier: the Prover can capture and double-check the hashes, ensuring no other information reaches the Verifier. Components drawn in orange represent components that the Verifier relies on to behave correctly, and components drawn in blue represent the same for the Prover.}
  \label{fig:macro}
\end{figure*}

\subsection{The Macroarchitecture (Placement of the Secure Gateway Device)}

\paragraph{Lifecycle of a packet going through the macroarchitecture.}
Traffic between the isolated AI data centre and the outside world (which includes company-internal peering metros) routes through a Wide Area Network (WAN) backbone router.\footnote{While early network security models relied on single, centralized hardware chokepoints, modern hyperscale architectures like Google's Espresso demonstrate that line-rate packet processing at the terabit scale must be decentralized~\cite{yap2017espresso}. Following this precedent, the SeGaDev would likely not be deployed as a single monolithic device. Instead, stateful operations (such as TCP sequence number translation and port randomisation) would be implemented as distributed software packet processors at a lower level, perhaps one per pod (still thousands of TPUs). This shards the state-table memory requirements across more devices, avoiding the extreme state-table memory needed in centralized hardware appliances at the cluster level.} This router marks the physical north-south boundary of the data centre network. Because modern distributed architectures encrypt all internal transit, packets passing through this router arrive as ciphertext --- typically encrypted via mutual TLS (mTLS) or proprietary internal protocols. Once inside the facility, this transport encryption is terminated at a frontend gateway. From there, the decrypted internal traffic is routed through the internal data centre fabric down to the individual compute units. This stage is where an SeGaDev could be placed to intercept plaintext traffic. We illustrate this in \Cref{fig:macro}, placing the device between ``TLS Termination'' and ``Internal Routing Fabric''. The decrypted packet traverses the SeGaDev. Protocol headers are scrubbed of degrees of freedom (port choices, initial sequence numbers (ISNs), timing) while application payloads are hashed for later verification (we detail the sanitization and hashing process happening inside this device in the next section). From there, the sanitised packet travels through the data centre-internal fabric to the individual compute unit (server, rack etc.). A compute unit transmitting results back is practically the same process in reverse, taking the same overall path.

This macroarchitecture relies on the following three assumptions:

\begin{itemize}
\item The facility is under physical monitoring, ensuring that any traffic in and out of the cluster needs to pass through the SeGaDev chokepoint.
\item Every bit of traffic sent to the Verifier (hashes only) can be passively observed by the Prover. This places the root of trust in physics: beam splitting via fused biconical taper (FBT) or planar lightwave circuit (PLC) splitters and optical isolators ensure the Prover can observe and double-check that hashes have been computed exactly as they expect, without being able to interfere with the Verifier's monitor leg. Timestamps can be computed by both parties upon receiving the hashes at practically the same time.
\end{itemize}

\Cref{fig:macro} represents a common case where transport encryption terminates before compute units. However, if the data centre additionally uses end-to-end encryption on internal links, the SeGaDev captures and hashes ciphertext while still scrubbing unencrypted protocol headers. The Prover stores session keys alongside their plaintext records, enabling a verification cluster to decrypt captured traffic after the sessions have concluded and the keys are not in active use.\footnote{For unfamiliar readers, TLS is the encryption between the outside world and the data centre. Unencrypted communication within data centres is not unusual~\cite{microsoft_ssl_overview}.}\textsuperscript{,}\footnote{If the Prover can freely choose encryption keys, this could be a steganography channel. One could require a deterministic key generation heuristic, but there may be security issues with such predictability. A solution may be to handle this via a coin flip protocol between the Prover and Verifier. A buffer of keys generated in the same way the SeGaDev generates ISNs.}

\subsection{The Microarchitecture (the Secure Gateway Device Internals)}
\label{sec:micro}

\begin{figure*}[htbp]
  \centering
  \includegraphics[width=\textwidth]{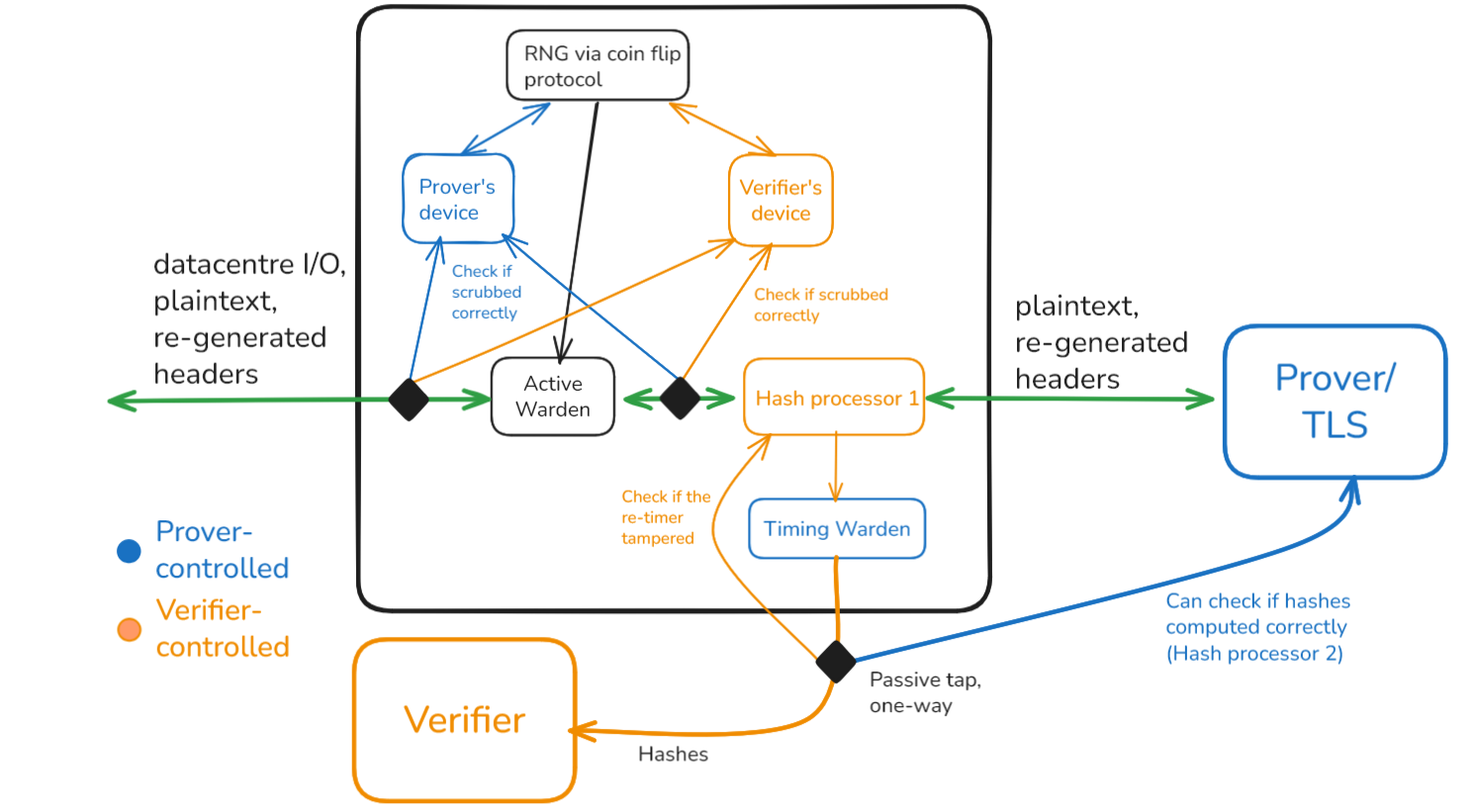}
  \caption{When north-south egress traffic from the data centre hits the SeGaDev, we first scrub non-reproducible fields with an Active Warden. Where necessary, we use a coin flip protocol as a source of randomness. Traffic leaving the Active Warden is hashed and forwarded. Components drawn in orange represent components that the Verifier relies on to behave correctly, and components drawn in blue represent the same for the Prover.}
  \label{fig:micro}
\end{figure*}

When a packet reaches the SeGaDev, it passes through two stages: sanitization and hashing. The hashes likewise pass through a warden for re-timing, eliminating potential covert communication to the Verifier. A key security feature used within the device is enabled by the passive (optical) splitters providing observability of untrusted components. In this architecture, no individual processor requires mutual trust. This property is valuable because establishing mutual trust in a single hardware component is difficult: \citet{ilhan2026verifiable} survey design- and fabrication-stage verification mechanisms and conclude that no single technique provides sufficient assurance against a high-capability adversary, with attack vectors including malicious third-party IP, dopant-level Trojans, and package-level die substitution, none of which are patchable post-deployment. By removing the need for such jointly trusted hardware, our design sidesteps this entire class of supply‑chain integrity problems. In \Cref{fig:micro}, we colour-code the minimum device trust assumptions. 
Transparent design and supply chains could also establish mutual trust in components, which would further add defense in depth, but our architecture does not rely on this.

\paragraph{Lifecycle of a packet going through the device.}

\paragraph{Coin flip protocol.} This stage prepares pseudorandom numbers for the warden to use for sanitizing packets, before these packets arrive. Standard network protocols contain unconstrained header fields such as ephemeral source ports, TCP Initial Sequence Numbers (ISNs), and session IDs. While replacing these values with a deterministic, post-hoc verifiable heuristic would eliminate the covert channel, doing so would fundamentally compromise the protocol's security --- predictability in fields like TCP ISNs or encryption keys exposes connections to sequence prediction attacks and session hijacking.

The Active Warden overwrites all such fields with pseudorandom values. If either party controlled this randomness, they could use it as a covert channel. This motivates introducing the coin-flip protocol, which can be summarised as follows. Both the Prover and the Verifier contribute their own number-generating devices for participating in the protocol. Each device first commits to a value (or a list of values), both reveal it after commitment, and the shared output is computed from both (e.g.\ using XOR on every bit). These values are pre-computed in bulk and consumed in strict FIFO order (`first in, first out'), so that no per-packet latency is introduced. Both devices can verify that the warden uses the correct value for each field by observing the sanitised output through their passive taps.

\paragraph{Sanitization (Active Warden).} Using the pre-generated random values (and heuristics where applicable, e.g.\ setting all padding bits to zero), the warden replaces all exploitable header fields and maintains bidirectional connection-tracking tables so that normal communication is preserved (see \Cref{app:ports}). This is known in literature as an Active Warden ~\citet{fisk2003eliminating} and ~\citet{xing2020netwarden}: a device that scrubs steganographic degrees of freedom from transit traffic. Prior implementations assume single-party trust in the device; our coin-flip protocol and passive observability eliminate this requirement.

An important part of sanitization is re-timing: even after header sanitization, the timing of packet transmission can carry information. An adversary could modulate inter-packet gaps or exploit analogue properties of the optical signal. The device buffers all packets and releases them on a fixed schedule at maximum link rate (PHY-Covert Channels~\cite{lee2014phy}). After this sanitization, traffic can be forwarded and leave the SeGaDev. Since hashes are computed only on the payloads, rather than also on the headers scrubbed by the Warden, the order of hashing and sanitization is not important and can differ between egress and ingress (in turn simplifying the design).

\paragraph{Hashing.} The packet is fed to a hash processor that computes a cryptographic commitment over a configurable group of packets referred to as the `hashing boundary' (see \Cref{app:verification}). These hashes --- and only these hashes --- are transmitted to the Verifier over a dedicated fibre link. A passive optical splitter on this link gives the Prover a copy, allowing them to independently verify that no information beyond hashes reaches the Verifier. The hash output itself passes through a separate, Prover-controlled re-timing stage (``Timing Warden'' in \Cref{fig:micro}) to prevent the hash processor from covertly communicating with the Verifier through timing or optical modulation. The Verifier's hash processor, in turn, can check if the Prover's re-timer tampered with the hashes, by also using a passive splitter on the re-timer's egress.

\section{Covert Channel Estimates}
\label{sec:covert}

We consider only covert channels through the SeGaDev in scope. Physical air-gapping of the monitored cluster ---apart from the monitored links--- is a requirement, but a separate challenge.

Making exfiltration of even single bits outright impossible is likely futile, as very low-bandwidth covert channels are inherently offense-dominant. The residual risk is not bulk data exfiltration (practically all meaningful computation output far exceeds what very low bandwidth channels can carry), but rather the combination of covert internal computation with minimal-bandwidth signalling of results. Defending against this requires complementary verification of compute utilisation (e.g.\ backend/east-west network taps, throughput accounting, numerical fingerprinting, etc.)~\cite{cankaya2025catching1}, which is outside the scope of the SeGaDev.

After the mitigations achieved via the Active Warden within the SeGaDev, we expect the following residual channels to remain:
\begin{itemize}
\item Remaining timing/order modulation
\item Steganography in ML outputs
\item Failure channel: not responding to select inputs and covering it up as ``hardware failures'' or ``software errors''
\end{itemize}

\subsection{Timing Modulation}

Re-timing in the SeGaDev can upper-bound the frequency of available timing modulation channels, depending on the buffer windows. With a 10~ms buffer window at 100~Gbps and $\approx$200-byte streaming packets, each window contains up to $\approx$625k packets. The attacker can modulate packet count per window, yielding up to $\log_2(625\text{k}) \approx 19$~bits/window. However, the dominant channel is packet ordering: unlike the single-flow threat model assumed in most covert channel literature~\cite{uttarwar2025delay,xing2020netwarden,park2023airgap}, in our setting the Prover controls all egress flows, making cross-connection interleaving a covert channel of potentially very high capacity, motivating the fixed-frame-count and canonical-ordering mitigations below.\footnote{A covert communication method where secret information is encoded by strategically manipulating the sequence in which packets from multiple independent data streams are merged onto a single outgoing network link.}

\paragraph{Mitigations:} By fixing a packet count per connection per window and giving a canonical ordering to the connections assigned by the Warden, we can make the packet stream for a given connection constant-rate and deterministic. The remaining covert bandwidth from timing modulation reduces to connection lifecycle events (opening/closing a given connection), reducing the covert bandwidth substantially.

\subsection{Steganography in ML Outputs}

Non-determinism of ML outputs offers steganographic degrees of freedom for the Prover, which have been studied in related literature. With replay verification using seed synchronization~\cite{rinberg2025verifying}, the residual covert bandwidth through ML inference output tokens is approximately 0.1~bits per token.\footnote{Making ``temperature'' sampling, as often used in generative AI, verifiable.}\textsuperscript{,}\footnote{~\cite{rinberg2025verifying} report $<0.5\%$ exfiltrateable information for Qwen3-30B-A3B at $10^{-4}$ FPR using three-tier FSSL + logit-rank classification. Their metric normalises against the theoretical maximum of $\log_2|V| \approx 17$~bits/token ($|V| = 151{,}936$ for Qwen3-30B-A3B). This gives $0.005 \times 17 \approx 0.1$~bits/token, i.e.\ a $\geq 200\times$ slowdown over the unverified channel.} Using this estimate, for a large inference cluster (200k GPUs at 2000 tokens/GPU/second) at maximum capacity, we calculate a covert egress bandwidth of approximately 40~Mbit/s, sufficient to exfiltrate a 1~TB model in under 3 days. Though this is the worst-case assumption of full inference utilization and the attacker having full control over all the outputs, it highlights the significance of the payload-steganography channel.

\paragraph{Mitigations:}
\begin{itemize}
\item \textbf{Eliminate the non-determinism.} The 0.1 bits/token residual exists because replay verification tolerates some deviations originating from floating-point non-associativity. Recent work shows this floor can be pushed toward zero. \citet{he2025nondeterminism} and \citet{cankaya2025catching1} show that within fixed conditions (same hardware, same kernels, same batch size) inference is already bitwise reproducible even for many performance-maximizing inference engines and settings. \citet{deepseekai2026v4} go further, open-sourcing end-to-end batch-invariant and deterministic kernels with supposedly minimal performance overhead in a frontier production MoE model, which demonstrates that determinism is not confined to research demonstrations and can be maintained at scale. 
The practical implication is that if the Prover is required to log their full metadata (batch composition, kernel versions, parallelism configuration, sampling seeds), the Verifier can either:
\begin{itemize}
    \item replay the computation on identical hardware without needing to trust it, by controlling information flow using network taps similar to the SeGaDev~\cite{cankaya2026catching2}; or
    \item emulate the Prover's hardware behaviour and the forward passes using deterministic software. This option would build on existing simulation work for individual tensor cores~\cite{khattak2025tensorcores} and expand the scope to whole inference pipelines.
\end{itemize}
\item \textbf{Active Warden scrubbing of hardware-related entropy.} Since only the trailing bits of token likelihood predictions are affected by hardware non-associativity, in principle it should be possible to add just enough noise to scrub any fluctuations at this order of magnitude, while still maintaining inference quality.\footnote{This may need to be interactive, though: autoregressive generation means that a sampled token becomes a prior for the next generation steps. This adds a round trip between a Warden and the compute cluster, and therefore some latency: the Warden round trip (send $\sim$50 filtered logits, receive selected token ID) adds microseconds per token (typical network latency for short-mid-range Ethernet). Against a $\sim$30~ms decode step, this is $<0.02\%$ overhead and falls entirely within the 10--100~ms re-timing jitter already applied for timing channel suppression. Bandwidth is $\sim$400~kB/s at 2000~tokens/s/GPU, trivial for wardens on 100~GbE links.}
\end{itemize}

\subsection{Failures and Errors}

The Prover controls when connections interrupt and hardware ``fails''. From the perspective of an external observer like the Verifier, a failure is the absence of a complete response. Public data suggests $\approx$50{,}000 GPU-hours per unexpected hardware failure~\cite{dubey2024llama3}, predicting $\approx$96 failures/day for a 200k GPU cluster.\footnote{From Meta's Llama~3 training~\cite{dubey2024llama3} on 16{,}384 H100 GPUs (419 interruptions over 54~days). This reflects hardware-level reliability and is used as an order-of-magnitude proxy; we found no inference-specific GPU MTBF (Mean Time Between Failures) data.} Assuming the Prover generates responses of up to 8000 tokens, each failure can encode at most $\log_2(8000) \approx 13$~bits in the choice of the interruption point. At baseline rates, this yields $\approx$1~kbit/day. Note that this is the worst-case assumption of full inference throughput across the entire data centre.

\paragraph{Mitigations:} Screen the data already hashed by the SeGaDev for suspicious failure rates. If the covert bandwidth is still considered intolerable, the Warden could inject ``pseudo-failures'' indistinguishable from real ones, as seen from the outside.

\section{Limitations and Future Work}
\label{sec:limitations}

Empirical validation of the setup is still required.\footnote{Still, we presented strong precedents for the individual components, and Active Wardens have been built to handle 100G line rates~\cite{xing2020netwarden}, which would in fact be over-provisioned for most long-range internet fibre connections. The per-connection state for TCP/UDP scrubbing (port translation tables, ISN offsets, connection phase tracking) requires approximately 200 bytes per connection (\Cref{app:ports}). Xing et al.\ demonstrated that programmable switch SRAM can sustain 200{,}000 concurrent connection states with comparable per-connection footprint~\cite{xing2020netwarden}, well exceeding the concurrent connection count expected on a data centre's north-south uplink. Although our Python-based simulation operated at $\sim$39~Mbps, production-grade software packet processors have been proven to operate at edge line rates. For instance, Google's ``Espresso'' architecture relies on host-based software processing to handle global peering traffic~\cite{yap2017espresso}, demonstrating that a single commodity server can process 37~Gbps with negligible CPU overhead.}

We expect the bill of materials of a somewhat representative device (FPGAs or NICs, transceivers, memory, custom mainboards and optical components) to cost less than the researcher's and engineer's time needed to build it. A small-scale demonstration can be as basic as a single inference server, FPGA-based SeGaDev and a verification server, while still being representative of larger scales (where one would add more of the same hardware).

The design of the SeGaDev, as well as the broader monitoring scheme, should undergo advanced red-teaming carried out by specialists in fields such as cybersecurity and physical security.

The whole architecture depends on the assumption that the SeGaDev is the only communication channel. Other covert channels exist beyond physical monitoring scope: RF emanation, power line signaling or even thermal channels from a facility that the Prover physically controls. The security problem is structurally similar to an insider threat, since the aim is to prevent secret communication between the Prover and their own hardware. We refer the reader to related works by~\citet{guri2024airgap} and~\citet{park2023airgap}.

Until solutions for imperfect reproducibility of ML outputs are found, steganography in output payloads dominates all other residual channels by orders of magnitude, making it the primary target for follow-on work.

We do not constrain bandwidth for covert communication \textit{into} the data centre. While the input data has many uncontrollable variables (media input, user prompts, etc.), input Wardens may re-write such information while preserving semantics. We leave this to future work. Threat modeling for very low outbound bandwidth --- but still impactful --- covert workloads in the data centre would further provide (or negate) motivation for such additional traffic sanitization.

Development and testing of network taps for backend traffic within the data centre can add defense in depth on top of air-gapping and combating covert channels through the SeGaDev.

\section{Conclusion}
\label{sec:conclusion}

In this work, we presented a design for cryptographically fingerprinting all traffic entering and leaving an AI data centre, with the goal of verifying that no undeclared workloads have been covertly exfiltrated from the facility. The core of our proposal is the SeGaDev: an Active Warden combined with a hashing pipeline, placed at the north-south network boundary of the cluster. By scrubbing steganographic degrees of freedom from protocol headers, re-timing packets to suppress analogue and timing side-channels, and hashing the sanitised traffic for later auditing, the device enables a cryptographic commitment to all egress data that the Prover can be later challenged to justify.

A key property of our design is that it removes the need for processors trusted by both parties. Instead, we achieve mutual verifiability via passive optical fibre splitters and coin-flip protocols for shared randomness. This grounds trust in the physics of beam splitting and optical isolation rather than in the integrity of any single device. Both the Prover and the Verifier can independently observe and check the device's behaviour without being able to interfere with each other's monitoring.

Our feasibility analysis suggests that the approach scales to clusters of 100k+ accelerators using commercially available hardware: FPGA-based network taps and hash processors, off-the-shelf optical components, and storage volumes that represent a negligible fraction of modern data centre costs. The individual building blocks --- traffic scrubbing, passive optical tapping, hardware hashing --- are already deployed in adversarial settings such as financial regulation and military networks.

For now, the core remaining challenge is that steganography in ML output payloads represents a significant residual covert channel, with a potential bandwidth of tens of Mbit/s for a fully utilised cluster. Closing this channel depends on progress in deterministic inference replay or active scrubbing of hardware-induced entropy in token likelihoods, both of which are active areas of research. Other residual channels --- timing modulation through connection lifecycle events and information encoded in faked hardware failures --- offer an adversarial Prover a much lower bandwidth. Additionally, the entire architecture rests on the assumption that the SeGaDev is the \textit{only} communication path out of the facility, which requires robust physical monitoring that we do not address.

We encourage the community to pursue a hardware demonstration of the SeGaDev, which we expect to be inexpensive relative to the human engineering effort involved. Given the pace of AI capability development and the growing interest in international agreements, verification infrastructure of this kind may be needed sooner than the policy frameworks it is meant to support.

\bibliographystyle{icml2026}
\bibliography{references}

\newpage
\appendix
\crefalias{section}{appendix}
\Crefname{appendix}{Appendix}{Appendices}
\crefname{appendix}{appendix}{appendices}
\onecolumn

\section{Prover's Verification Mechanism}
\label{app:verification}

While the exact detail of the verification cluster, as well as the verification pipeline itself, are out of scope for this paper, we now briefly describe one possible workflow. The fundamental challenge is as follows: given a hash value chosen by the Verifier out of the stored pool of hashes, how can the Prover reliably and efficiently demonstrate that they possess a record which hashes to this value?\footnote{As a reminder, in this work we do not consider the problem of verifying that declared workloads are compliant with regulations, only the problem of verifying that no other undeclared workloads have been run.}

A simple idea is to maintain a traditional `hash table' --- a dictionary of all plaintext LLM inputs, the corresponding plaintext outputs and their hash values. Since the Prover already controls the egress from the data centre, they can easily register how many data packets (e.g., Ethernet frames) each output is broken into. Furthermore, since the SeGaDev specification would be fully open-source, they also know the `hashing boundary' --- how many frames are included in one hash. This allows the Prover to have a causal association between inputs, outputs and the corresponding sequence of their hashes. The hash values flow back to the Prover through a passive network tap placed on the connection which carries them from the SeGaDev to the verification cluster (see \Cref{fig:macro}).

The number of Ethernet frames within each hashing boundary corresponds to the fundamental `verification unit' and is arbitrary. It should strike a balance between storage requirements and recomputation time. For example, an extreme choice would be to hash every single Ethernet frame, leading to the highest possible number of hashes in storage. Since in this scenario one hash corresponds to a very small number of output tokens, this minimises the number of outputs that need to be recomputed in the verification facility. On the other hand, we could hash together all Ethernet frames that pass the SeGaDev in a month, which would lead to only one hash per month in storage, but would require one month of recomputation during verification.

The Prover needs to maintain knowledge of which inputs and parts of the corresponding outputs were included in each hash. For example, assume that the hashing window is 150 Ethernet frames, two input prompts generate 120 and 70 tokens, respectively, and one output token is streamed out via a single Ethernet frame. The Prover must then associate the following information with the hash values $h_1$ and $h_2$:

\begin{verbatim}
{
    h1: [(prompt_1, output_1, 1, 120),
         (prompt_2, output_2, 1, 30)],
    h2: [(prompt_2, output_2, 31, 70), ...],
    ...
}
\end{verbatim}

In this way, when the Verifier demands that hash $h_1$ be reproduced, the Prover can immediately look up the fact that prompts 1 and 2 must be sent to the verification facility (alongside the corresponding outputs and their indices).\footnote{Note that in general, these could be batched prompts or some other type of input. We used single prompts for simplicity in this illustrative example.} This information is necessary for the facility to faithfully reproduce the exact stream of 150 Ethernet frames that should be hashed and compared to $h_1$. If this new hash matches $h_1$, the Verifier is satisfied that the Prover's cluster must have originally carried out the same workload. If this verification pipeline is employed enough times, the Verifier can gain confidence that the Prover did not execute hidden workloads --- because they would not have had the time and free computational resources in their cluster to execute such workloads in addition to the ones that were verified post hoc.

\section{Storage Requirements}
\label{app:storage}

To demonstrate the feasibility of our proposed verification pipeline, we now estimate the upper bound of the volume of hashes and plaintext data that would need to be stored by the Verifier and the Prover respectively. First, let us assume that the data centre under monitoring contains 100k GPUs of the NVIDIA Hopper generation. Each such GPU is assumed to be capable of a 2000 token/s inference throughput, meaning that the total throughput of the cluster is $2 \times 10^8$ tokens/s.\footnote{Deepseek's inference setup for their v3/R1 models is reported to achieve a token throughput of $\sim$14.8k tokens per H800 node (8 GPUs, decode unit) per second.} Next, we need to calculate how many bytes of data this token volume corresponds to. For this, we assume that the tokens are wrapped in OpenAI's streaming API template of the following format:

\begin{verbatim}
{
  "id": "chatcmpl-123",
  "object": "chat.completion.chunk",
  "created": 1694268190,
  "model": "gpt-4o-mini",
  "system_fingerprint": "fp_44709d6fcb",
  "choices": [
    {
      "index": 0,
      "delta": {
        "content": "Hello"
      },
      "logprobs": null,
      "finish_reason": null
    }
  ]
}
\end{verbatim}

Importantly, note that this template includes a single response token only. This is because the streaming API aims to deliver generated tokens to the user as quickly as possible, rather than waiting for the full generation to finish before sending the payload. The `one-token-per-json' scenario represents the most challenging configuration for our methodology, as this will lead to the highest possible number of hash computations.

Furthermore, we operate under an (unrealistic) assumption that each such template is allocated to one Ethernet frame. Therefore, the total number of Ethernet frames streamed out of the cluster is simply $2 \times 10^8$ per second, the same as number tokens. The template carries around 200 ASCII characters, corresponding to 200 bytes (+ headers) --- well below the Maximum Transmission Unit (MTU) of the Ethernet protocol (1500 bytes). We then calculate the SHA-256 hash of each such frame, leading to $2 \times 10^8 \times 24 \times 3600 \times 365 = 6.3 \times 10^{15}$ hash computations performed in a year. The corresponding hash volume is $6.3 \times 10^{15} \times 256 / 8 = 2.0 \times 10^{17}$~bytes $= 200$~PB.

Additionally, the Prover needs to maintain a year-long record of all plaintext data transmitted from their cluster to the outside world. Each generated token is assumed to be wrapped in the template above. Since one ASCII character is stored by exactly 1 byte, 200 bytes are required to stream one token out of the cluster.\footnote{The Prover must store the complete streaming API response templates rather than the bare output tokens. This is because the hashes computed by the SeGaDev are taken over the Ethernet frames as transmitted, which include the full JSON payload with all metadata fields.} On an annual scale, this corresponds to $2 \times 10^8 \times 200 \times 24 \times 3600 \times 365 = 1.26 \times 10^{18}$~bytes $\sim 1{,}260$~PB. Note that we do not separately estimate the storage requirements for input data (prompts), as they are negligible in comparison.\footnote{While input prompts are typically longer than individual output tokens (especially in multi-turn conversations), they are stored as single objects rather than in per-token JSON templates. Even under a pessimistic assumption that input prompts are 20x longer than outputs in terms of the raw token count, the per-token storage overhead from the streaming format dominates by a factor of around $200/20 = 10$.}

The cost of storing such a large amount of data is small ($<1\%$) compared to the costs of AI data centres at the scale considered here: a typical price for a 28~TB HDD hard drive is around 500\euro, thus the full cost amounts to $1{,}260~\text{PB}/28~\text{TB} \times 500\text{\euro} \sim 22{,}500{,}000$~\euro.\footnote{As of early 2026.}

Note that all these estimates represent a pessimistic upper bound that arises from \textit{extremely} inefficient data management. When using non-streaming APIs, multiple tokens are included as part of one chat completion template. Further, multiple such templates could be transmitted as part of a single Ethernet frame, as long as the total payload size does not exceed the standard Ethernet MTU of 1500 bytes. Storing API JSON files for each token is also likely unnecessary when they can be reconstructed from tokens and heuristics. We also did not include data compression at all. Finally, we do not need to compute the hash of each Ethernet frame. Instead, the `fundamental unit' of verification could be an arbitrarily long chain of frames. All these modifications could reduce storage by multiple orders of magnitude.

\section{Preventing Dictionary Attacks}
\label{app:dictionary}

The Prover might be worried that the Verifier can deduce the plaintext content of the messages passing through the SeGaDev by guessing which input hashes to the observed hash values. Deducing this content by inverting the hash is not possible --- hash functions are by design irreversible. However, an adversarial Verifier could be tempted to use so-called dictionary attacks. In other words, they could pre-compute the hashes of the most common plausible inputs and simply compare them to the observed ones. The feasibility of this approach is determined by the expected size of the input space of the hash function. For example, if there can only ever be 10 possible inputs, then it is trivial to pre-compute their 10 corresponding hashes and later infer the input that has produced the hash coming out of the SeGaDev.\footnote{Note that this might not actually be an issue at all, since in our method, the hashed content also includes metadata such as timestamps of the API response. If the adversarial Verifier does not know these, then dictionary attacks are not possible in the first place. However, we do see this threat vector as important, as in the future AI outputs might be wrapped in standardized templates that do not include unpredictable metadata entries. This could be, for example, to reduce storage requirements on Prover's side (see \Cref{app:storage}). In such cases, dictionary attacks become relevant.}

Unfortunately for the Prover, the input space of their API responses is indeed somewhat limited: these can only be tokens generated by the LLM. These in turn are constrained by the size of the LLM vocabulary, typically around 200{,}000 tokens. Thus, in the scenario presented above --- where only one output token is inserted into the streaming API template, this template corresponds to one Ethernet frame and later one hash --- the attacker could pre-compute the hashes of all tokens inserted into these templates. They would then simply perform the lookup in a hash table with 200{,}000 values and immediately infer the right token. Thus, it would be trivial for the Verifier to guess all the output tokens from the streaming API responses.

To calculate how many tokens need to be inserted into the output template to prevent such attacks via combinatorial complexity, let us consider how many hash computations a motivated attacker could perform in a month. Specialised chips can achieve throughputs of 200~TH/s (terahashes per second). We use this Bitcoin-mining figure as a worst-case upper bound. Against the SeGaDev's longer inputs, an attacker would need a custom ASIC and would likely achieve 1–2 orders of magnitude lower throughput in practice, but we conservatively assume the Bitcoin rate. Within a month, this corresponds to $5.2 \times 10^{20}$ hashes computed. On the other hand, a sequence of $n$ tokens out of a 200k vocabulary can lead to $(2 \times 10^5)^n$ distinct combinations. Thus, the attacker is guaranteed to guess the sequence corresponding to a given hash if the sequence is shorter than $\log_{2 \times 10^5}(5.2 \times 10^{20}) \sim 3.9$ tokens long. Running multiple specialised chips in parallel will not help the attacker much --- with 1000 chips, the token number increases to 4.5.

To address this, we recommend that the streaming API template should enclose at least 5 tokens in order to prevent brute-force hash guessing by the Verifier. However, this estimate assumes that the attacker needs to try all possible $n$-gram combinations of tokens from the vocabulary. This is an unrealistic assumption, as the tokens streamed out of the cluster will almost always be semantically meaningful --- sequences of tokens that form sentences, code snippets, equations, or similar. Thus, the search space is greatly reduced by noting that only a small fraction of the vocabulary is likely to follow the previous token. Assuming that only 1000 most likely tokens need to be checked, our `effective vocabulary' is reduced to 1000. Then, the calculation becomes $\log_{1000}(5.2 \times 10^{20}) \sim 7$.

In other words, any semantically meaningful sequence of fewer than 7 tokens could be brute-forced by the Verifier in less than a month with a single hash-crunching chip.\footnote{Once again, we note that this discussion considers worst-case assumptions. In reality, an adversarial Verifier would probably need to somehow know in advance which hashes to attack --- presumably, only specific AI outputs would be of interest to them (for example, those carrying sensitive intellectual property), not any randomly selected sequences of outputs. Additionally, a month-long attack period is something that only the most motivated of attackers are likely to attempt.} Thus, to prevent this, the streaming API template could wrap at least 10 tokens. This should have a negligible effect on user experience --- instead of seeing tokens appear on their screen one at a time, the user will receive them 10 at a time. Alternatively (or additionally), the hashes could simply be computed on more than one JSON template or more than one Ethernet frame. For example, if the hashing boundary is set every 100 Ethernet frames, the whole problem disappears even with 1 streaming token per JSON template.

\section{Precedents, Available Hardware and Feasibility}
\label{app:feasibility}

While our proposal is conceptual for now, we emphasize that all of the individual components for the SeGaDev are either established and field-tested (passive optical fibre splitters, coin flip protocols, hash functions with timestamps~\cite{digicert_rfc3161}) or built from quickly reconfigurable commercial hardware.

For information processing, Field-Programmable Gate Arrays (FPGAs) are a suitable choice for two reasons. First, they can be adapted to any specialised application after manufacturing. Second, they are inherently more secure against supply chain attacks, since an attacker cannot anticipate the circuit design in advance.\footnote{With some limited exceptions. For example, there are fixed I/O locations.} For these reasons, FPGAs find use in a range of adversarial, high-stakes environments, most notably military networks (Infodas SDoT Secure Network Card, Infodas SDoT Security Gateway), high-frequency trading, and financial law enforcement. FPGA network cards and network taps are commercially available from multiple vendors (e.g.\ BittWare IA-780i, Arista 7130 Series). These systems are deployed at scale, operate at line rate (10--400~Gb/s), introduce negligible latency, and are relied upon for regulatory compliance where evidence integrity is legally contested (e.g., CAT NMS / Reg NMS~\cite{sec_rule613}, MiFID II~\cite{ateam_mifid2}).\footnote{A specific example is Metamako/Arista Networks, whose 7130 FPGA taps are used for precision timestamping~\cite{arista_timestamps}. The German stock exchange (Deutsche B\"orse) publicly disclosed deployment of dozens of such devices in a case study~\cite{arista_deutscheboerse}. Their threat model in trading networks is explicit: traders attempt to hide behaviour~\cite{cftc_panther}; regulators require complete, lossless, timestamped records~\cite{arista_timestamps}.}

Traffic scrubbing~\cite{xing2020netwarden}, traffic normalization, and re-timing devices~\cite{uttarwar2025delay} against timing and analogue side channels have all been built and tested at line rates exceeding 100~Gb/s.

We now give a brief feasibility assessment from the bandwidth and storage perspective. In \Cref{app:storage}, we estimate that a 100k GPU AI cluster streaming 200M tokens per second corresponds to an egress API bandwidth of $\sim$40~GB/s. The SeGaDev, even under such extreme assumptions, can use off-the-shelf hardware. Modern hashing algorithms are efficiently parallelizable on mid-range FPGAs.\footnote{The SHA-3 256 hashing function is a natural choice for hardware hashing at high bandwidth, achieving $\sim$34~Gbps per core in $\sim$1,375 slices ($\sim$5,500 lookup tables, aka.\ LUTs) on a Virtex-7 FPGA. Twelve parallel cores sustain $\sim$410~Gbps hashing in $\sim$66,000 LUTs, roughly 13\% of a commodity Kintex UltraScale+ KU15P (522,720 LUTs, $\sim$\$4,000 or lower at volume).} The storage requirements for both the Prover and the Verifier are insignificant for any realistic data management solution (see \Cref{app:storage}).

Inference clusters, network taps and verification clusters can be quickly demonstrated on small scales, while still being almost fully representative of large-scale deployments: a single server or rack to simulate the Prover's and Verifier's clusters respectively. Scaling the system becomes a matter of adding more of the same hardware. At the scale of real-world deployments, the main technical challenge then shifts to physical security and red-teaming.

For the sub-scale demonstration, we expect the bill of materials to be insignificant compared to the cost of the engineering and security red-teaming staff required for a pilot demonstration.

\section{Eliminating Steganography Needs Connection Tables}
\label{app:ports}

Preventing steganographic communication between the data centre and the outside world introduces the need to eliminate certain degrees of freedom from the protocol metadata. To illustrate this more concretely with an example, in this Appendix we justify why source and destination ports need to be explicitly handled by the Active Warden.

Network protocols grant both parties control over certain fields. In UDP, the initiating party (client) selects its source port, whilst the responding party (server) must use the destination port specified in the request as its source port for the response. Without intervention, either party could encode information by selecting these ports strategically.

Consider a simple example: an external client sending API requests to a data centre. By choosing source port 5000 versus 5001, the client transmits one bit of information per connection. At 1000 requests per second, this yields 125~Bps or $\sim$10~MB/day of covert channel capacity. Similarly, when the data centre initiates outbound connections (e.g., fetching training data), it could encode information in its choice of the ephemeral source port (typically ranging from 32768 to 60999, providing $\log_2(28232) \sim 15$~bits per connection). The Active Warden must eliminate both channels.

Two separate connection-tracking tables are required, because connections can be initiated from either direction and the security logic differs in each case:

\begin{itemize}
\item \textbf{External-initiated connections} (e.g.\ API requests to the data centre) allow an external client to encode information in their choice of the client source port. The Warden must replace this with a random value, then maintain a state table to translate responses back to the original port.
\item \textbf{data centre-initiated connections} (e.g.\ fetching external datasets) allow the data centre operator to encode information in their choice of ephemeral source port. Again, the Warden must replace this with a random value and maintain a reverse translation state.
\end{itemize}

Maintaining separate tables for each direction simplifies the translation logic, as the scrubbed field and the party being defended differ between the two cases. Both tables have keys characterised by the full connection 4-tuple: (source IP, source port, destination IP, destination port).

Each state-tracking table has two distinct security functions during the connection lifecycle. First, during the request phase, the table performs port randomisation: the initiating party's source port (client port) is replaced with a value from the RNG, eliminating their ability to encode information. Simultaneously, the table records the target server port from the request's destination field. During the response phase, the table performs enforcement and back-translation: it verifies the responding party uses the correct server source port (which prevents the server from encoding information through port choice), and it translates the response destination port from the RNG value back to the original client port. Overall, this ensures that the response is delivered to the correct client port and does not carry information steganographically.

Similar translation tables need to be maintained wherever parts of the relevant network protocol offer degrees of freedom that could be used for covert communication. This could introduce significant programmatic complexity for more complicated protocols like TCP, which are stateful. In TCP, the Active Warden must additionally perform sequence number translation on every single packet, and in both directions. More precisely, during the so-called `TCP handshake', the client's Initial Sequence Number (ISN) is replaced with a random number. Thus, the Warden must record the offset between the original and scrubbed ISN, so that all subsequent packets in this connection have their outbound sequence numbers incremented by this consistent offset. Similarly, the corresponding inbound ACK numbers in this connection need to be decremented by the same offset. This is necessary to maintain consistency in both the original and scrubbed sequences.

Overall, the Active Warden must track connection state to handle the handshake phase, data transfer phase, and connection termination. Each TCP connection might require around 200 bytes of state-tracking data. For a data centre with $O(1000)$ concurrent connections processing millions of packets per second, the Active Warden must implement efficient table lookups to avoid slowdowns in serving this traffic to customers.

\newpage
\section{Glossary}
\label{app:glossary}

For convenience, this appendix collects acronyms and terms used in the
paper.

\subsection*{Acronyms}

\begin{description}
  \setlength{\itemsep}{2pt}
  \item[FBT] Fused Biconical Taper. An optical splitter formed by fusing
        two fibres together and tapering them so that light couples between
        them.
  \item[FPGA] Field-Programmable Gate Array. A reconfigurable digital logic
        chip whose internal connections are programmed after manufacture.
  \item[FSSL] Fixed-Seed Sampling Likelihood
        \citep{rinberg2025verifying}. Given a sampling seed and a claimed
        token, the probability the token could have been produced honestly
        under benign numerical noise.
  \item[HBM] High Bandwidth Memory. 3D-stacked DRAM placed adjacent to a
        processor die for very high memory bandwidth.
  \item[ISN] Initial Sequence Number. The 32-bit starting sequence number
        negotiated during a TCP handshake.
  \item[LUT] Look-Up Table. The basic logic primitive of an FPGA; implements
        arbitrary Boolean functions of a few inputs.
  \item[MTU] Maximum Transmission Unit. The largest payload size that fits
        in a single frame of a given link layer (1500\,B for standard
        Ethernet).
  \item[mTLS] Mutual TLS. A TLS handshake in which both client and server
        authenticate to each other with certificates.
  \item[NIC] Network Interface Card. The hardware connecting a host to a
        network.
  \item[OSI] Open Systems Interconnection. The seven-layer reference model
        used to describe networking functionality, from the physical layer
        up to the application layer.
  \item[PHY] Physical Layer. Layer~1 of the OSI model; the electrical,
        optical, or RF signaling on the wire.
  \item[PLC] Planar Lightwave Circuit. A silica-on-silicon optical
        waveguide chip used as a splitter or combiner.
  \item[RNG] Random Number Generator.
  \item[SRAM] Static Random-Access Memory. Fast on-chip memory that holds
        state without periodic refresh.
\end{description}

\subsection*{Terms}

\begin{description}
  \setlength{\itemsep}{2pt}
  \item[Active Warden] A network element that actively rewrites traffic
        (rather than only inspecting it) to remove covert channels
        \citep{anderson1998limits}.
  \item[Coin flip protocol] A commit-reveal protocol in which both parties
        commit to a value, both reveal, and the joint output (e.g.,
        bitwise XOR) is used as shared randomness; neither party can bias
        the result unilaterally.
  \item[Hashing boundary] The unit of network traffic over which a single
        cryptographic commitment is computed by the SeGaDev (e.g.\ one
        packet, a packet group, or a time window).
  \item[Prover / Verifier] The two roles in the verification protocol. The
        Prover is the operator of the AI compute facility being inspected;
        the Verifier is the external party checking compliance.
  \item[Re-timing] Buffering packets and releasing them on a fixed
        schedule to remove timing-based covert channels.
  \item[SeGaDev] Secure Gateway Device. The verification device proposed
        in this paper.
\end{description}

\newpage

\end{document}